\documentclass[twocolumn,secnumarabic,amssymb, nobibnotes, aps, pre]{revtex4-1}
\usepackage{graphicx}
\usepackage{dcolumn}
\usepackage{bm}
\usepackage[usenames]{color}
\usepackage[toc]{appendix}

\usepackage[all]{xy}
\usepackage{amsmath}
\usepackage{natbib}
\usepackage{longtable}
\usepackage{amsmath}

\def \bea{\begin{eqnarray}}
\def \eea{\end{eqnarray}}
\def \be{\begin{equation}}
\def \ee{\end{equation}}

\def\({\left(} \def\){\right)}
\def \tx{\widetilde{x}}
\def \ty0{\widetilde{y}_0}

\begin{document}
\bibliographystyle{apsrev4-1}

\title{Stable Swarming Using Adaptive Long-range Interactions}
\author{Dan Gorbonos and Nir S. Gov}
\email[Corresponding author:]{nir.gov@weizmann.ac.il}
\affiliation{Department of Chemical Physics, The Weizmann Institute of Science, Rehovot 76100, Israel}

\begin{abstract}
Sensory mechanisms in biology, from cells to humans, have the property of adaptivity, whereby the response produced by the sensor is adapted to the overall amplitude of the signal; reducing the sensitivity in the presence of strong stimulus, while increasing it when it is weak. This property is inherently energy consuming and a manifestation of the non-equilibrium nature of living organisms. We explore here how adaptivity affects the effective forces that organisms feel due to others in the context of a uniform swarm, both in two and three dimensions. The interactions between the individuals are taken to be attractive and long-range, of power-law form. We find that the effects of adaptivity inside the swarm are dramatic, where the effective forces decrease (or remain constant) with increasing swarm density. Linear stability analysis demonstrates how this property prevents collapse (Jeans instability), when the forces are adaptive. Adaptivity therefore endows swarms with a natural mechanism for self-stabilization.
\end{abstract}

\maketitle

\setcounter{equation}{0}  
\setcounter{figure}{0}

\section{Introduction} In recent years there is an intense interest in the physics of collective migration of organisms, being examples for out-of-equilibrium active matter \cite{vicsek2012collective,marchetti2013hydrodynamics}. The description of these systems is mostly in terms of active agents that interact with each other through a variety of short and long range interactions. The source of these interactions depends on the sensory systems that the biological entities posses, such as visual, chemical or acoustic. Many sensory mechanisms, in a variety of organisms, are subject to a modification due to adaptivity~\cite{wark2007sensory}. This property should affect most significantly the collective behavior when the interactions are long-range, typically in power-law decay form.

Adaptivity is the process whereby the sensitivity of the sensory mechanisms drops when there is a strong input. This feature prevents both damage and saturation of the sensory organs. Exact adaptation means that the steady-state output is independent of the steady-state level of input. This is part of a fold-change detection mechanism~\cite{shoval2010}, which is ubiquitous in nature, and involves a response whose entire shape, including amplitude and duration, depends only on fold change and not on the absolute levels of the input. The response usually includes a strong transient part, followed by adaptation which determines the steady-state at longer-times.
Famous examples include cell receptors~\cite{friedlander2009adaptive} and molecular signaling tasks in cells~\cite{goentoro2009evidence,cohen2009dynamics,tyson2003sniffers,brandman2008feedback}, as well as collective behaviour which is based on chemical signaling~\cite{torney2009context} such as in the case of bacteria~\cite{shklarsh2011smart}. Another important example on the cellular level is chemotaxis~\cite{tu2008modeling,barkal1997robustness,tindall2008overview}. Finally, there is adaptation also in senses such as vision, touch and hearing in animals and humans~\cite{gardner2000principles,smirnakis1997adaptation}.

In this paper we explore the effects of adaptivity within swarms, where individuals interact by attractive long-range interactions. These interactions, of power-law form, occur when organisms are responsive to mechanical stimuli, such as hydrodynamic flows~\cite{drescher2010direct,zaid2016analytical}, elastic deformations~\cite{schwarz2002elastic,schwarz2013physics} or acoustic interactions~\cite{previous}. The addition of adaptivity is motivated by our recent work on acoustic interactions in midge swarms~\cite{previous}. Adaptive alignment interactions in flocks were previously studied in~\cite{Eitan1,Eitan2}. Our broader motivation is to study adaptive forces as a novel class of physical interactions. We assume here that the adaptivity is fast compared to the motions of the organisms, such that the sensory system is all the time in its steady-state, long-time response regime.

\section{Adaptive power-law forces.} We first define the adaptive form of the forces that arise from power-law stimuli. We write the total (directional) intensity of the signal that an organism $i$ receives from $N$ other organisms as the vector sum of the stimuli due to all the pair interactions $\vec{S}_{ij}$:
 \be
 \vec{S}_{i}=\sum_{j=1}^{N}\vec{S}_{ij},\: \vec{S}_{ij}=\frac{C_n}{r_{ij}^{n+1}}\,(\vec{r}_i-\vec{r}_j) \label{s}
 \ee
where $r_{ij}\equiv |\vec{r}_{i}-\vec{r}_{j}|$ is the distance between the emitter and the receiver, and we define $C_n$ as a constant with dimensions of $mass \cdot length^{n+1}/time^{2}$, so that the stimulus and the effective force have the same dimensions. $C_n$ is negative for attractive forces. We assume here that the sensory organ is able to discern the direction from which the signal is received, while the amplitude is proportional to the intensity. In addition, we ignore angular dependencies which exist for higher multipoles of emitted signals, if the emitters/receivers rotate on a fast time-scale or are oriented randomly.

We assume here that all the organisms are identical, and treat the response of the organism to the stimulus as an effective force acting on it~\cite{previous}. Due to the adaptivity mechanism, this effective force (response function) $\vec{F}_{i}$ does not depend on the total amplitude of the stimulus. This is achieved by a rescaling of the stimuli by the inverse of its total amplitude (sum over all the received intensities according to Eq.~\ref{s})
 \be
 N_{\mbox{\scriptsize tot},i}\equiv\sum_{j=1}^{N}|\vec{S}_{ij}|, \:\vec{F}_{i}\propto\frac{\vec{S}_{i}}{N_{\mbox{\scriptsize tot},i}}=\frac{|C_n|}{R_{\mbox{\scriptsize ad}}^n}\cdot\frac{\sum_{j=1}^{N}\vec{S}_{ij}}{N_{\mbox{\scriptsize tot},i}},\label{perfect}
 \ee
where $R_{\mbox{\scriptsize ad}}$ is a constant with dimensions of $length$. This normalization first appeared in the context of alignment interactions~\cite{Eitan1,Eitan2}.
With this ansatz the effective force is sensitive only to the relative difference in the strength of the sources. In other words, the force $ \vec{F}_{i}$ is invariant under the rescaling $ \vec{S}_{ij} \rightarrow \alpha \vec{S}_{ij}$, which is exactly the definition of the fold change detection mechanism~\cite{shoval2010}.

Input amplitudes in most sensory systems can vary by many orders of magnitudes~\cite{shoval2010}. At low signal levels, the cost of sensing the input received might exceed its benefit. On the other end, at high input levels, saturation might affect the system. Thus, adaptivity takes place only in a finite range of input stimuli. For this purpose we can introduce into the model a finite range in the form of a characteristic length $R_{\mbox{\scriptsize ad}}$ over which adaptivity occurs, in the form
 \be
  \vec{F}_{i}=\frac{\sum_{j=1}^{N}\vec{S}_{ij}}{1+|C_{n}|^{-1}\,R_{\mbox{\scriptsize ad}}^n
  \sum_{j=1}^{N}|\vec{S}_{ij}|}.\label{Fadapt}
 \ee
 Since $|\vec{S}_{ij}| \sim r_{ij}^{-n}$, when the distances between pairs are large $r_{ij}\gg N^{\frac{1}{n}}R_{\mbox{\scriptsize ad}}$ the adaptivity does not play a role. On the other hand when $r_{ij}<N^{\frac{1}{n}}\,R_{\mbox{\scriptsize ad}}$ adaptivity is strong and Eq. (\ref{perfect}) is a good approximation, which we term ``perfect adaptivity''. Most of this paper we will work in this regime. Note that the effective force between two organisms under perfect adaptivity becomes a constant, independent of the separation between them.

We will consider integer values of $n\geq2$, while for our original motivation of acoustic (density) waves~\cite{previous} the dominant contributions are for even $n$ (see appendix~\ref{AppA}). The interaction that we consider here is purely attractive between organisms, while in real swarms there is an additional short-range repulsion in order to avoid collision. We omit this detail in our analysis. In addition, individuals in a living swarm have a noisy component to their motion~\cite{previous}, but the average acceleration felt by each agent is described by the effective power-law force that we consider here.

\section{The effective force in a spherical swarm} We consider here the case of a three-dimensional swarm, where the organisms move in a uniform spherical swarm of radius $R_s$ (given in appendix~\ref{AppB} in cylindrical coordinates) and we calculate the effective force within it. Note that since we deal with uniform density we take the continuum analogues of the sums in the equations above. We define a local coordinate system (Fig.~\ref{cylindricalcoordinates}) around a given point ($r=0$, $z=z_0$), to write the stimulus $\vec{S}_{(n)}=S_{(n)}\hat{z}$ and total amplitude integrals (Eqs.~(\ref{TotalStimulus1})-(\ref{TotalAmplitude1})).
\begin{figure}[tb]
\centering
\includegraphics[width=\linewidth]{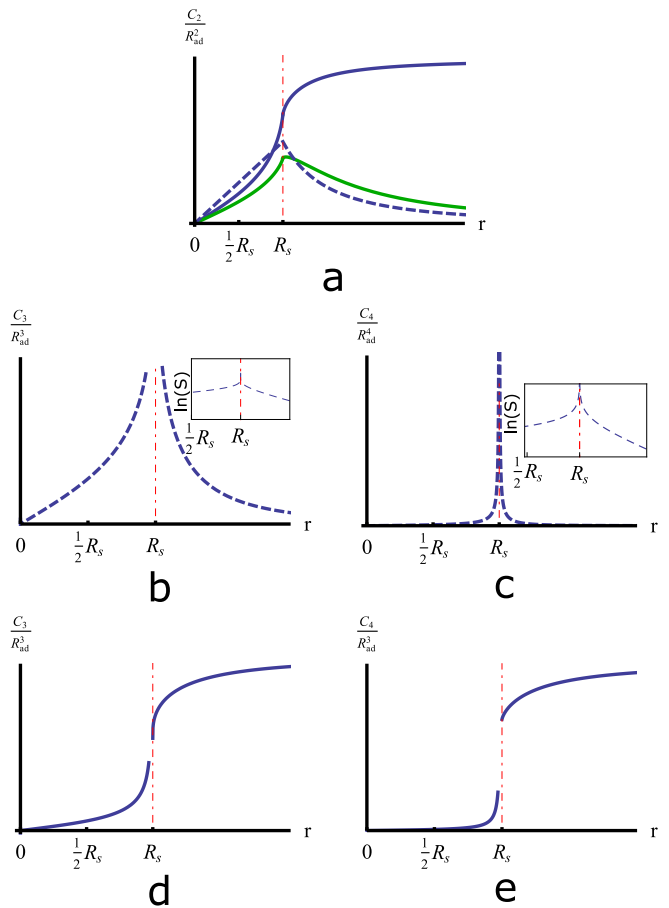}
\caption{\label{theforce}  (a) The effective force (thick line) and the stimulus (dashed line) for $n=2$. Both are functions of $z_0$ and in units of $C_2/R_{\mbox{\scriptsize ad}}^2$. The red dashed line indicates the radius of the swarm at $r=R_s$. The effective force is plotted in the ``perfect adaptivity'' regime where $R_s \ll N^{\frac{1}{n}}R_{\mbox{\scriptsize ad}}$. The green line is the same effective force but with $R_{\mbox{\scriptsize ad}}\sim\sqrt{3}/\sqrt{4\,\pi\,\rho\,R_s}$.  (b,c) The stimulus for $n=3,4$, respectively. (d,e) The effective force for $n=3,4$, respectively. The gap around $r=R_s$ is due to the cut-off that we use here $r_c=0.03\,R_s$. The insets show the same plots in log-log scale.  }
\end{figure}
Outside the swarm ($z_0>R_s$) these integrals are well behaved. For any $n\geq 3$ we have to introduce a cut-off since the integrals diverge for $z_0\leq R_s$. This apparent divergence is a manifestation of the fact that the force diverges with higher powers than the mass contribution, which increases as the volume $r^3$. The case of $n=2$ is special since we do not encounter any divergencies. The $n=2$ interaction was termed ``adaptive gravity'' and introduced as a model for interactions in insect swarms~\cite{previous}. For this special case we get
\be
S_{(2)}(z_0)=\begin{cases}
    \frac{4\,\pi\,\rho\,C_2}{3}z_0, & z_0<R_{s}\\
    \\
    \frac{4\,\pi\,C_2\,R_{s}^3\,\rho}{3\,z_0^2}, & R_{s}<z_0.
  \end{cases}\label{S2}
\ee
and
\be
N_{\mbox{\scriptsize tot},(2)}(z_0)=\pi\,\rho\,|C_2|\,[2R_{s}-\frac{(R_s^{2}-z_0^{2})}{z_0}\ln\(\frac{|R_s-z_0|}{R_s+z_0}\)].\label{Ntot2}
\ee


The effective force (substituting Eqs.~(\ref{S2})-(\ref{Ntot2}) into Eq.~(\ref{perfect})) and the stimulus for the $n=2$ case are plotted in Fig.~\ref{theforce}a. Note that while in regular gravity (the bare stimulus in this case) the attractive force decays outside of the swarm, for perfect adaptivity the effective force does not decay. However, in reality the adaptivity does diminish beyond the length-scale of $N^{\frac{1}{n}}R_{\mbox{\scriptsize ad}}$ (Eq.~(\ref{Fadapt})), and the force does decay far from the swarm.  For a uniform spherical symmetric swarm with a radius of $R_s$, where $N =4\,\pi \rho R_s^3/3$, the adaptivity length $R_{\mbox{\scriptsize ad}}$ which gives a range on the order of the size of the swarm, is given by (Eq.~(\ref{Fadapt})): $R_{\mbox{\scriptsize ad}}\sim 3^{\frac{1}{n}}\,R_s^{1-\frac{3}{n}}/\(4\,\pi\,\rho\)^{\frac{1}{n}}.$ In the case of $n=2$ we get $R_{\mbox{\scriptsize ad}}\sim \sqrt{3}/\sqrt{4\,\pi\,\rho\,R_s}$, and the effective force for this value of $R_{\mbox{\scriptsize ad}}$ is plotted in Fig.~\ref{theforce}a.

For $n>2$ we introduce a cut-off at a distance $r_c=\rho^{-\frac{1}{3}}$ (Fig.~\ref{cylindricalcoordinates}). Instead of spherical coordinates we use cylindrical coordinates, and arrive at the following analytical expressions
\vspace{2 mm}
 \begin{widetext}
 \be
 \label{TotalStimulus}
 S_{(n)}(z_0)
=\begin{cases}
\frac{2\,\pi\,\rho\,C_n}{(n-1)(n-3)(n-5)z_0^2}\left[(R_{s}+z_0)^{3-n}\left(R_{s}^2+(n-3)R_s\,z_0+z_0^2\right)-|R_{s}-z_0|^{3-n}\left(R_{s}^2-(n-3)R_s\,z_0+z_0^2\right)\right] & n\neq 3,5 \\
 \pi\,\rho\,C_3 \left[\frac{1}{2}\(\frac{R_s^2}{z_0^2}+1\)\ln\(\left|\frac{R_{s}+z_0}{R_{s}-z_0}\right|\)-\frac{R_{s}}{z_0}\right] & n=3\\
 \pi\,\rho\,C_5 \left[\frac{R_s\(R_s^2+z_0^2\)}{2\,z_0\,(R_s^2-z_0^2)^2}+\frac{1}{4\,z_0^2}\ln\(\left|\frac{R_{s}-z_0}{R_{s}+z_0}\right|\)\right] & n=5
 \end{cases}
 \ee

\be
N_{\mbox{\scriptsize tot}, (n)}(z_0)=\begin{cases} \left|\frac{2\,\pi\,\rho\,C_n}{(n-4)(n-3)(n-2)z_0}\lbrace \frac{(R_{s}+z_0)^3}{|R_s+z_0|^n}\left[(n-3)R_s+z_0\right]-\frac{(R_{s}-z_0)^3}{|R_s-z_0|^n}\left[(n-3)R_s-z_0\right]\rbrace+\frac{4\,\pi\,C_n}{(n-3)}\,\rho^{\frac{n}{3}}\,\Theta\(R_s-z_0\)\right| \label{TotalAmplitude} & n>4 \\
2\,\pi\,\rho\,|C_3|\left[\ln(R_s^2-z_0^2)+\frac{2}{3}\ln\rho\right]\Theta\(R_s-z_0\)+2\,\pi\,\rho\,|C_3|\left[\ln(\frac{z_0+R_s}{z_0-R_s})-\frac{2\,R_s}{z_0}\right]\Theta\(z_0-R_s\) & n=3\\
\left|\frac{C_4}{C_2}\cdot\frac{N_{\mbox{\scriptsize tot},(2)}}{ R_s^{2}-z_0^{2} }-4\,\pi\,C_4 \rho^{\frac{4}{3}} \,\Theta\(R_s-z_0\)\right| & n=4\\
 \end{cases}
\ee
\end{widetext}
where $\Theta(x)$ is the Heaviside step function. For $z_0<R_s$ we find that $N_{\mbox{\scriptsize tot}, (n)}$ is a function of the cut-off radius $r_c$, while
$S_{(n)}$ is independent of the cut-off. Similar behavior is also found in two-dimensions (see appendix~\ref{AppF}).

The stimulus $S_{(n)}$ does not diverge inside the sphere, for the following reason: For any point inside the sphere there are sources that give divergent contributions in all directions and eventually cancel out to give a finite stimulus in the limit of $r_c\rightarrow0$. On the boundary, however, all the sources are on one side and they add up to give a divergence for $n\geq3$.  When considering the finite distance between organisms in the swarm, of order $r_c$, the stimulus near the swarm edge ($z_0=R_s\pm r_c$) is also finite. We use these analytical results to plot the stimuli and effective forces for the cases of $n=3,4$ (Eqs.~(\ref{F3})-(\ref{F4})), as shown in Fig.~\ref{theforce}b-e. It is interesting to note that: $F_{(4)}=F_{(2)}$ for $z_0>R_s$ (Eq.~(\ref{F4})), and that simpler expressions for the effective force are obtained for $n\ge6$, as fractions of polynomials in $z_0$ (see appendix~\ref{AppD}).

Close to the swarm center ($z_0 \ll R_s$) we get from Eq.~(\ref{TotalStimulus})
\be
S_{(n)}(z_0)=\frac{4\,\pi\,\rho\,C_n\,z_0}{3\,R_{s}^{n-2}}+\mathcal{O}(z_0^3),\label{sn3dcenter}
\ee
and from Eq.~(\ref{TotalAmplitude}) for $n>2$
\be
N_{\mbox{\scriptsize tot}, (n)}=\begin{cases}
   \frac{2\,\pi\,\rho\,|C_3|}{3}\ln(R_s^6\,\rho^2)+\mathcal{O}(z_0^2), & n=3\\
    \\
    \frac{4\,\pi\,\rho\,|C_n|}{|n-3|}\(\rho^{\frac{n-3}{3}}-\frac{1}{R_s^{n-3}}\)+\mathcal{O}(z_0^2), & n>3.
  \end{cases} \label{Ntotcenter}
\ee
Therefore the effective force in the center is
\be
F_{(n)}(z_0)=\begin{cases}
\frac{C_2}{R_{\mbox{\scriptsize ad}}^2}\cdot\frac{z_0}{3\,R_s}+\mathcal{O}(z_0^2), & n=2\\
\\
\frac{C_3}{R_{\mbox{\scriptsize ad}}^3}\cdot\frac{2\,z_0}{R_s \ln(R_s^6\,\rho^2)}+\mathcal{O}(z_0^2), & n=3\\
\\
\frac{C_n}{R_{\mbox{\scriptsize ad}}^n}\cdot\frac{|n-3|z_0}{3\,R_s}\cdot\frac{1}{\rho^{\frac{n-3}{3}}\,R_s^{n-3}-1}+\mathcal{O}(z_0^2). & n>3
 \end{cases}\label{Fcenter}
\ee

We find that for all $n\ge 3$ the effective force near the center decreases with increasing density of the sources, while the signal $S_{(n)}$ is linear in the density (Eq.~(\ref{sn3dcenter})). This behavior, for all $n\ge3$ (or $n\ge2$ in two dimensions, Eqs.~(\ref{Sn2dcenter})-(\ref{F2dcenter}), is due to the fact that the intensity increases with increasing density faster than the geometric decrease in the number of neighbors. For the special case of $n=2$ we find that the force near the center is independent of the density.

\section{Dynamic instability} So far we calculated the average forces within a static swarm. We can now carry out a linear stability analysis to determine the size of stable swarms. Such an analysis, for the case of gravity, leads to the famous criterion for Jeans instability that is usually related to the collapse of interstellar gas clouds and subsequently star formation~\cite{jeans,binney2011galactic}. We extend here a heuristic derivation of the Jeans instability to include adaptive power-law interactions and in appendix~\ref{AppH} we give a more rigorous derivation following~\cite{binney2011galactic}.
Note that the Jeans instability criterion is calculated for an underdamped system, since an overdamped system does not allow for a collapse instability since the force and therefore the velocity diminish as the objects approach the center.

Consider an infinite homogeneous swarm with uniform density $\rho_0$. Due to local fluctuations a region whose length scale is $L$ becomes denser $\rho_0 \rightarrow \rho_0+\delta \rho$. The overdense region will collapse if the random velocities of the organisms are not large enough to carry them out of the region before the collapse due to the attractive force can occur. The typical escape time (due to random velocities) from the overdense region is
$t_{\mbox{\scriptsize esc}}=L/\sigma_v$ where $\sigma_v=\sqrt{\langle v^2\rangle}$ is the root mean square velocity. The r.m.s. velocity can arise from thermal motion, from the chaotic motion due to the attractive forces themselves (influenced by the initial conditions), as well as due to (and maybe dominated by) the noisy active propulsion forces of the living agents (such as for the midges in a swarm \cite{kelley2013}). The criterion for instability is $t_{\mbox{\scriptsize esc}}>t_{\mbox{\scriptsize col}}$ where $t_{\mbox{\scriptsize col}}$ is the typical time to collapse. For a linear restoring force of the form $\vec{F}=-K\,\vec{r}$ the typical time for collapse is the time of the order of the period
$t_{\mbox{\scriptsize col}}\sim 2\,\pi/\sqrt{K}$ (for a unit mass). In the case of pure gravity $K=4\,\pi\,|C_2|\,\rho/3$, and therefore the critical density of the region is
\be
\rho_{\mbox{\scriptsize Jeans}}\sim\frac{\pi\,\langle v^2\rangle}{L^2\,|C_2|}. \label{crit1}
\ee
When the density is higher than the critical density, $\rho>\rho_{\mbox{\scriptsize Jeans}}$, there is an unstable mode and the region will undergo a collapse. Using Eq.~(\ref{sn3dcenter}) we can generalize this result for $n\geq 2$  power-law force close to the center of the swarm ($K=4\,\pi\,|C_n|\,\rho/(3\,R_s^{n-2})$), taking $L \sim R_s$, we get
\be
\rho_{\mbox{\scriptsize Jeans}}^{(n)}\sim\frac{\pi\,L^{n-4}\,\langle v^2\rangle}{|C_n|}.
\ee
Adding adaptivity, the effective force is given in Eq.~(\ref{Fcenter}), and then the same argument gives for $n=2$ the following criterion for instability:
\be
\langle v^2\rangle\lesssim\frac{L\,|C_2|}{\pi^2\,R_{\mbox{\scriptsize ad}}^2},
\ee
which is independent of the density. Therefore in swarms with an adaptive gravity force we cannot reach collapse just by increasing the density, as opposed to the pure gravitational case.

For $n\geq 3$ we get the opposite inequality for the density. The critical density
\be \label{criticalDens}
\rho_{\mbox{\scriptsize Jeans,ad}}^{(n)}=\begin{cases}
\frac{1}{L^3}e^{\frac{ \mathcal{N} |C_3|\,L}{\pi^2\langle v^2\rangle\,R_{\mbox{\scriptsize ad}}^3}} & n=3\\
\\
\frac{1}{L^3}\(1+\frac{\mathcal{N} L\,|C_n|\,|n-3|}{R_{\mbox{\scriptsize ad}}^n\,\pi^2\langle v^2\rangle}\)^{\frac{3}{n-3}}. & n>3
 \end{cases}
\ee
is now the maximal density for instability (where $\mathcal{N}$ stands for a number of order one), and the swarm becomes stable for any higher density. Thus the adaptivity produces a self-stabilization mechanism that also contributes to the cohesiveness of the swarm. When the density is low, the swarm is unstable and tends to collapse until it reaches the density that is given by Eq.~(\ref{criticalDens}). In other words starting from low density, we see that adaptive forces tend to pick a particular density $\rho_{\mbox{\scriptsize Jeans,ad}}^{(n)}$ in which the swarm is stable.

\section{Conclusion} We have explored here the effects of adaptivity on the effective forces acting on organisms within and outside uniform swarms, both in two and three dimensions, when the interactions between them is of power-law form. We find that the effects of adaptivity are dramatic: outside the swarm the interactions do not decay with distance from the swarm, as long as adaptivity is effective. Inside the swarm, the effective forces decrease (or remain constant) with increasing swarm density, unlike the stimulus or regular non-adaptive interactions which increase linearly with the density of the sources. This opposite dependence of the central attraction inside the swarm on the density of sources helps to protect adaptive swarms from collapse: when the density of an adaptive swarm increases, the attractive effective force towards the center decreases, which allows the random motion of the organisms to dilate the swarm and reduce its density. As the density decreases the attractive force increases until a balance is reached. In non-adaptive interaction, such as gravity, clusters have a tendency to undergo instability and collapse, since the attractive force increases with the density (Jeans instability~\cite{jeans}). Adaptivity therefore endows swarms with a natural mechanism for stabilization. These results should apply to a wide range of biological systems.

\appendix

\section{Even multipoles for acoustic interactions}
\label{AppA}

In the case of acoustic interactions the dominant contributions come from even values of $n$. We assume that the stimulus is proportional to the intensity of the lond-range interactions that are detected by the work that they do on the sensory organs. This work is proportional to the energy flux $\vec{q}$ of the waves~\cite{landau1987fluid}:
 \be
 \vec{q}=\delta p \, \vec{v}, \label{energyflux}
 \ee
where $\delta p$ is the variation of the pressure from its background value and $\vec{v}$ is the velocity of the particles that are displaced by the wave. Let us assume that there is one dominant multipole in the pressure variation
\be
\delta p\sim\frac{1}{r^{l}}\cos(\vec{k}\cdot\vec{r}-\omega t +\phi_0)+\mathcal{O}\(\frac{1}{r^{l+1}}\),
\ee
where $l$ is the multipole number (a positive integer), $\vec{r}$ is the distance from source, $\vec{k}$ is the wave number, $\omega$ is the frequency and $\phi_0$ is the phase. For a traveling plane wave we get~\cite{landau1987fluid}
\be
\delta p=c\,\rho_0\,v,
\ee
where $c$ is the speed of sound and $\rho_0$ is the density of the transmission medium. Substituting into Eq.~(\ref{energyflux}) and taking the time average we get
that the leading term of the energy flux is
\be
\langle q \rangle \sim \frac{1}{r^{2\,l}},
\ee
namely we get even powers of the distance.

\section{The total stimulus $S_{(n)}$ and the total amplitude $N_{\mbox{\scriptsize tot}}$ in three dimensions}

\label{AppB}

We will use cylindrical coordinates $(r,z,\varphi)$ and calculate the field at  ($r=0$, $z=z_0$) without loss of generality (the point $A$ in Fig.~\ref{cylindricalcoordinates}). The symmetry of the problem implies that the field is along the $z$ axis. For power-law interactions of the form in Eq.~(\ref{s}), the contribution of a point at $(r',z')$ to the stimulus at $(0,z_0)$ is
\[
\frac{C_n\,\rho}{(r'^2+(z'-z_0)^2)^{\frac{n}{2}}},
\]
and the angle is
\[\cos\varphi=\frac{z'-z_0}{\sqrt{r'^2+(z'-z_0)^2}}.\]
Hence the total stimulus at $z_0$ is

\be
S_{(n)}(z_0)=2\,\pi\,C_n\,\rho\int_{-R_s}^{R_s}\!\!\!dz'\int_{0}^{\sqrt{R_{s}^2-z'^2}}\!\!\!\!\!r'dr'\,\frac{z_0-z'}{\left[r'^2+(z'-z_0)^2\right]^{\frac{n+1}{2}}} \label{TotalStimulus1}\ee
and the normalization factor, $N_{\mbox{\scriptsize tot}}$, which is given by the total amplitude of the signal at the point $(0,z_0)$ is
\be
N_{\mbox{\scriptsize tot}, (n)}(z_0)=\left| 2\,\pi\rho\,C_{n}\int_{-R_s}^{R_s}\!\!\!dz'\int_{0}^{\sqrt{R_{s}^2-z'^2}}\!\!\!\!\!\frac{r'dr'}{\left[r'^2+(z'-z_0)^2\right]^{\frac{n}{2}}} \right|.
\label{TotalAmplitude1}
\ee

Since there seems to be a divergence at $(0,z_0)$, we can rewrite the integrals above in spherical coordinates $(\tilde{r},\varphi,\chi)$ centered at $(0,z_0)$ and see that for $n>2$ the integrals diverge.

\bea \label{stim3d}
\!\!\!\!\!\!\!\!\!S_{(n)}&=& 2\,\pi\rho\,C_{n}\!\!\!\int_{0}^{\pi}\!\!\!\sin(\varphi)\,\cos(\varphi)\,d\varphi\!\!\int_{0}^{\tilde{R_s}(\varphi)}\!\!\!\!\!\!d\tilde{r}\,\tilde{r}^{2-n}, \nonumber\\
N_{\mbox{\scriptsize tot}, (n)}&=&\left| 2\,\pi\rho\,C_{n}\!\!\!\int_{0}^{\pi}\!\!\!\sin(\varphi)d\varphi\!\!\int_{0}^{\tilde{R_s}(\varphi)}\!\!\!\!\!\!d\tilde{r}\,\tilde{r}^{2-n}\right|.
\eea

 \section{The calculation in three dimensions with a cut-off}

\label{AppC}

\begin{figure}[tb]
\centering
\includegraphics[width=0.6\linewidth]{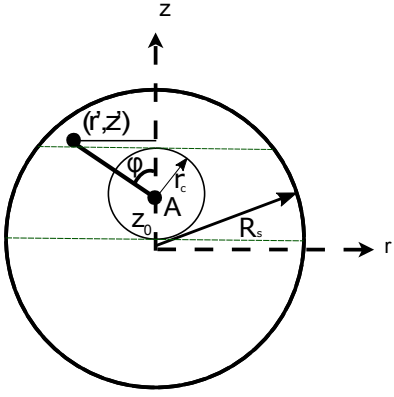}
\caption{\label{cylindricalcoordinates}  The cylindrical coordinates $(r,z)$ (and $\varphi$) that we use for the calculation of the effective force at a point A in a uniform-density spherical swarm. The cut-off at the radius $r_c$ from $A$ is indicated as well. The green lines separate the three regions of integration that we used.}
\end{figure}

 We divide the integrals of the stimulus and the amplitude (which are given without cut-offs in Eqs.~(\ref{TotalStimulus1})-(\ref{TotalAmplitude1})) into three sectional volumes as indicated in Fig.~\ref{cylindricalcoordinates}.
\begin{widetext}
\bea
S_{(n),r_c}(z_0)&=&\int_{-R_s}^{z_0-r_c}\!\!\!dz'\int_{0}^{\sqrt{R_{s}^2-z'^2}}\!\!\!dr'I_{S}+\int_{z_0-r_c}^{z_0+r_c}\!\!\!dz'\int_{\sqrt{r_{c}^2-(z'-z_0)^2}}^{\sqrt{R_{s}^2-z'^2}}\!\!\!dr'I_{S}+\int_{z_0+r_c}^{R_s}\!\!\!dz'\int_{0}^{\sqrt{R_{s}^2-z'^2}}\!\!\!dr'I_{S},\\
N_{\mbox{\scriptsize tot},(n)}^{r_c}&=&\left|\int_{-R_s}^{z_0-r_c}\!\!\!dz'\int_{0}^{\sqrt{R_{s}^2-z'^2}}\!\!\!dr'I_{N}+\int_{z_0-r_c}^{z_0+r_c}\!\!\!dz'\int_{\sqrt{r_{c}^2-(z'-z_0)^2}}^{\sqrt{R_{s}^2-z'^2}}\!\!\!dr'I_{N}+\int_{z_0+r_c}^{R_s}\!\!\!dz'\int_{0}^{\sqrt{R_{s}^2-z'^2}}\!\!\!dr'I_{N}\right|,
\eea
\end{widetext}

where

\bea
I_{S}&\equiv&2\,\pi\rho\,C_{n}\,\rho\frac{r'\,(z_0-z')}{\left[r'^2+(z'-z_0)^2\right]^{\frac{n+1}{2}}}, \\
I_{N}&\equiv&2\,\pi\rho\,C_{n}\,\rho\frac{r'}{\left[r'^2+(z'-z_0)^2\right]^{\frac{n}{2}}}.
\eea

\section{The total stimulus and the effective force for various values of $n$ in three dimensions}

\label{AppD}
For $n=3$ the stimulus $S_{(3)}$ is given in Eq.~(\ref{TotalStimulus}) and the total amplitude can be derived from Eq.~(\ref{TotalAmplitude}). Substituting into the definition of the effective force, as given in Eq.~(\ref{perfect}) we get:
\be \label{F3}
F_{(3)}(z_0)=\begin{cases}
    \frac{3C_3}{ 4\,z_0^2\,R_{\mbox{\scriptsize ad}}^3}\frac{\left[(z_0^2+R_{s}^2)\ln\(\frac{R_{s}+z_0}{R_s-z_0}\)-2\,R_s\,z_0\right]}{\ln\left[\rho^2(R_s^2-z_0^2)^3\right]}, & z_0<R_{s}\\
    \\
     \frac{C_3}{ R_{\mbox{\scriptsize ad}}^3}\left[\frac{1}{4}\(\frac{R_s^2}{z_0^2}+1\)\right.\\
     \left.-\frac{R_s^3}{2\,z_0^2\(2\,R_s+z_0\ln\(\frac{z_0-R_s}{z_0+R_s}\)\)}\right], & R_{s}<z_0.
  \end{cases}
\ee

Next, we calculate the case of $n=4$. The stimulus is given by
\be
S_{(4)}(z_0)=\frac{C_4}{C_2}\cdot\frac{S_{(2)}(z_0)}{|R_s^2-z_0^2|}.
\ee
It turns out that for $z_0>R_s$
\bea
\!\!\!\!\!\!\!\!\!F_{(4)}(z_0)&=&\frac{C_4}{R_{\mbox{\scriptsize ad}}^4}\cdot\frac{S_{(4)}(z_0)}{N_{\mbox{\scriptsize tot}, (4)}}=\nonumber\\
&=&\frac{C_2}{R_{\mbox{\scriptsize ad}}^2}\cdot\frac{S_{(2)}(z_0)}{|R_s^2-z_0^2|}\Big{/} \frac{N_{\mbox{\scriptsize tot},(2)}}{\left | R_s^{2}-z_0^{2} \right|}=F_{(2)}(z_0),
\eea
where we identify the constants $C_2=C_4\cdot R_{\mbox{\scriptsize ad}}^2$.
Hence one cannot distinguish between the two power-law interactions $n=2,4$ when there is adaptivity outside the swarm. The full effective force for $n=4$ is given by
\be
F_{(4)}(z_0)=\begin{cases}
    \frac{F_{(2)}(z_0)}{\left|1-4\,\pi\,C_2\,\rho^{\frac{4}{3}}\,(R_s^2-z_0^2)\,N_{\mbox{\scriptsize tot}, (2)}^{-1}\right|}, & z_0<R_{s}\\
    \\
     F_{(2)}(z_0), & R_{s}<z_0.
  \end{cases}\label{F4}
\ee

For $n=6$ the total stimulus is
\be
S_{(6)}(z_0)=\begin{cases}
  \frac{4\,\pi\,\rho\,C_6\,z_0\,(5\,R_s^2-z_0^2)}{15\,(R_{s}^2-z_0^2)^3}, & z_0<R_{s}\\
  \\
   \frac{4\,\pi\,\rho\,C_6\,R_s^3\,(5\,z_0^2-R_s^2)}{15\,z_0^2\,(z_0^2-R_{s}^2)^3}, & R_{s}<z_0.
  \end{cases}
\ee
and the total amplitude is
\be
N_{\mbox{\scriptsize tot},(6)}(z_0)=\left|\frac{4\,\pi\,\rho\,C_6\,R_s^3}{3(R_s^2-z_0^2)^3}-\frac{4\,\pi\,C_6}{3}\,\rho^2\,\Theta(R_s-z_0)\right|,
\ee
so that the effective force is:
\be
F_{(6)}(z_0)=\begin{cases}
    \frac{C_6}{ R_{\mbox{\scriptsize ad}}^6}\cdot \frac{z_0\,(5\,R_s^2-z_0^2)}{5\,\left|R_s^3-(R_s^2-z_0^2)^3\,\rho\right|} , & z_0<R_{s}\\
    \\
     \frac{C_6}{R_{\mbox{\scriptsize ad}}^6}\cdot (1-\frac{R_s^2}{5\,z_0^2}), & R_{s}<z_0.
  \end{cases}
\ee

Higher even powers $n>6$ give expressions for the effective force that are fractions of polynomials in $z_0$ as in the $n=6$ case.

\section{Asymptotic behavior of the stimuli and the effective force}

\label{AppE}
From Eqs.~(\ref{TotalStimulus},\ref{TotalAmplitude}) we find that far away from the swarm (assuming perfect adaptivity) $z_0\gg R_s$:
\be
 S_{(n)}(z_0),N_{\mbox{\scriptsize tot}, (n)}(z_0) \rightarrow \frac{4\,\pi\,\rho\,C_n}{3\,z_0^{n-3}}\(\frac{z_0}{R_s}\)^3+\mathcal{O}((\frac{z_0}{R_s})^5),
\ee
and then according to Eq.~(\ref{perfect}) the effective force goes asymptotically to the following constant value
\be
F_{(n)}(z_0) \rightarrow \frac{C_n}{R_{\mbox{\scriptsize ad}}^n}+\mathcal{O}((\frac{z_0}{R_s})^5).
\ee

\section{The effective force in a two dimensional circular swarm.}

\label{AppF}

We consider here the case of a two-dimensional swarm, where the organisms move on a flat plane. The swarm in this case is a uniform disk of radius $R_s$ with a density $\sigma$ of organisms. We use cartesian coordinates $(x,y)$ and calculate the field at $(x=0,y=y_0)$ without loss of generality (Fig.~\ref{coordinates2}a). The symmetry of the problem implies that the net field is along the $y$ axis: $\vec{S}_{(n)}=S_{(n)}\hat{y}$ (where $n$ is the power in Eq.~(\ref{s})). For power-law interactions of the form in Eq.~(\ref{s}), the contribution of a point at $(x',y')$ to the stimulus at $(0,y_0)$ is
\[
S(x',y')=C_n\sigma\frac{1}{(x'^2+(y'-y_0)^2)^{\frac{n}{2}}},
\]
and the angle is (see Fig.~\ref{coordinates2}a)
\[\cos\varphi=\frac{y'-y_0}{\sqrt{x'^2+(y'-y_0)^2}}.\]
Hence the total stimulus and amplitude at $y_0$ for $n\geq 2$ are
\begin{widetext}
\bea
S_{(n)}(y_0)&=&C_n\,\sigma\int_{-R_s}^{R_s}\!\!\!dx'\int_{-\sqrt{R_{s}^2-x'^2}}^{\sqrt{R_{s}^2-x'^2}}\!\!\!\!\!dy'\,\frac{y_0-y'}{\left[x'^2+(y'-y_0)^2\right]^{\frac{n+1}{2}}}
 \label{TotalStimulus2d}\\
 &=& \frac{C_n\,\sigma}{n-1}\int_{-R_s}^{R_s}\!\!\!dx'\left[\frac{1}{(R_{s}^2-2\,y_0\,\sqrt{R_{s}^2-x'^2}+y_0^2)^{\frac{n-1}{2}}}-\frac{1}{(R_{s}^2+2\,y_0\,\sqrt{R_{s}^2-x'^2}+y_0^2)^{\frac{n-1}{2}}}\right].\nonumber
\eea

\begin{figure}[tb]
\centering
\includegraphics[width=0.6\linewidth]{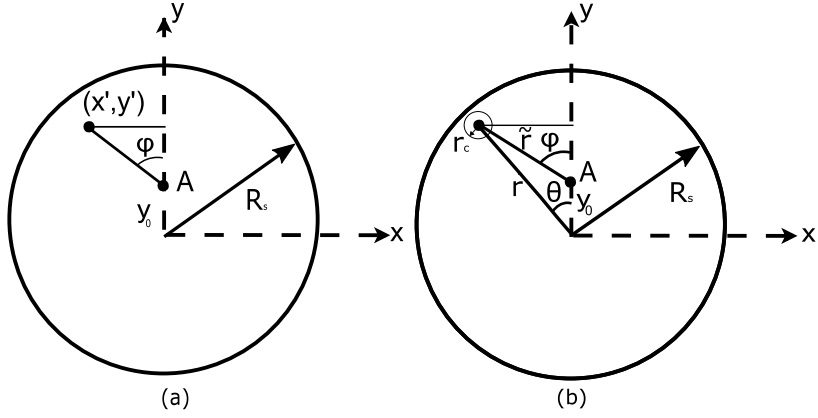}
\caption{ (a) The cartesian coordinates $(x,y)$ that we use for the calculation of the effective force at a point A in a uniform-density circular disk.
(b) The new polar coordinates $(\tilde{r},\varphi)$}.
\label{coordinates2}
\end{figure}

\bea
N_{\mbox{\scriptsize tot}, (n)}&=&\left| C_n\,\sigma\int_{-R_s}^{R_s}\!\!\!dx'\int_{-\sqrt{R_{s}^2-x'^2}}^{\sqrt{R_{s}^2-x'^2}}\,\frac{dy'}{\left[x'^2+(y'-y_0)^2\right]^{\frac{n}{2}}}\right|=\nonumber\\
&=&\left|\frac{C_{n}\sigma}{R_s^{n-2}}\int_{-1}^{1}\frac{d\tx}{\tx^2}\left[\frac{\sqrt{1-\tx^2}+\ty0}{(1+2\,\ty0\,\sqrt{1-\tx^2}+\ty0^2)^{\frac{n}{2}-1}}\,F^{+}_{n}(\tx)+\frac{\sqrt{1-\tx^2}-\ty0}{(1-2\,\ty0\,\sqrt{1-\tx^2}+\ty0^2)^{\frac{n}{2}-1}}\,F^{-}_{n}(\tx)\right]\right|.
\label{Ntot2d}\eea
\end{widetext}
where $\tx\equiv x/ R_s$ and $\ty0 \equiv y_0 / R_s$ are dimensionless variables, and
\[
F^{\pm}_{n}(\tx)\equiv _{2}\!F_{1}
\left(1,\frac{3-n}{2},\frac{3}{2}\,  ; -\frac{(\ty0\pm \sqrt{1-\tx^2})^2}{\tx^2}
\right).\]

$_{2}\!F_{1}(\alpha,\beta,\gamma;x)$ is a hypergeometric
function defined by
\[_{2}\!F_{1}(\alpha,\beta,\gamma;z)=\sum_{n=0}^{\infty}\frac{(\alpha)_{n}\,(\beta)_{n}}{(\gamma)_{n}}\,z^{n}\,,\]
where $(\alpha)_{n} \equiv \Gamma(\alpha+n)/\Gamma(\alpha)$.

$N_{\mbox{\scriptsize tot}, (n)}$ diverges when $|\ty0| \leq 1$ for $n \geq 2$.
It happens due to the singular point at $(0,y_0)$. As an example we show explicitly this divergence for $n=2,4$ below. The divergence for any value of $n \geq 2$ will be be easy to see when we write the integral in~(\ref{Ntot2d}) in polar coordinates.

The origin of this divergence comes from the transition to the continuum limit. It disappears when we take into account the fact that the surface density is finite and introduce a cut-off at a distance of $r_c=\sigma^{-\frac{1}{2}}$ from the point $(0,y_0)$ (when $y_0<R_s$). For this purpose let us introduce polar coordinates $(\tilde{r},\varphi)$ centered at $(0,y_0)$ (see Fig.~\ref{coordinates2}b).

Then the normalization factor, $N_{\mbox{\scriptsize tot}}$, at the point $(0,y_0)$ is
\be \label{cutoff}
N_{\mbox{\scriptsize tot}, (n)}=\left| C_n\,\sigma\int_{0}^{2\,\pi}\!\!\!d\varphi\int_{r_c}^{\tilde{R_s}(\varphi)}\!\!\!d\tilde{r}\,\tilde{r}^{1-n}\right|,
\ee
and we see that this integral diverges for $n \geq 2$ when $r_c \rightarrow 0$.
$\tilde{R_s}(\varphi)$ is computed from the geometry of the new radial coordinates, as given in Fig.~\ref{coordinates2}b:

\be \label{cosineT}
\tilde{R_s}=\sqrt{R_{s}^2+y_0^2-2\,R_{s}\,y_0\,\cos \theta},
\ee
where $\theta$ is measured at the center of the disk.
From the sine rule
\[\frac{\tilde{R_s}}{\sin{\theta}}=\frac{R_s}{\sin{\varphi}}\]
it follows that
\[\cos \theta =\ty0\,\sin^2\varphi\pm \sqrt{\cos^2{\varphi}(1-\ty0^2\,\sin^2\varphi)}.
\]
Substituting into Eq.~($\ref{cosineT}$) we get for $n=2$:
\be \label{RegularizedNtot2d}
N_{\mbox{\scriptsize tot}, (2)}=\left| C_2\,\sigma\left(\int_{-\frac{\pi}{2}}^{\frac{\pi}{2}}\!\!\!d\varphi\ln\tilde{R}_s^{+}+\int_{\frac{\pi}{2}}^{\frac{3\pi}{2}}\!\!\!d\varphi\ln\tilde{R}_s^{-}+\pi\ln\sigma\right)\right|,
\ee
where
\[\frac{\tilde{R}_s^{\pm}}{R_s}\equiv\(1+\ty0^2\pm2\ty0\left[\sqrt{\cos^2{\varphi}(1-\ty0^2\,\sin^2\varphi)}\mp\ty0\,\sin^2\varphi\right]\)^{\frac{1}{2}}.\]
For $n>2$ we have:
\be \label{RegularizedNtot3d}
N_{\mbox{\scriptsize tot}, (n)}=\left|\frac{C_n\,\sigma}{n-2}\left[ \int_{-\frac{\pi}{2}}^{\frac{\pi}{2}}\!\!\!\frac{d\varphi}{(\tilde{R}_s^{+})^{n-2}}+ \int_{\frac{\pi}{2}}^{\frac{3\pi}{2}}\!\!\!\frac{d\varphi}{(\tilde{R}_s^{-})^{n-2}}-2\,\pi\,\sigma^{\frac{n-2}{2}}
\right]\right|.
\ee

\begin{figure}[tb]
\centering
\includegraphics[width=\linewidth]{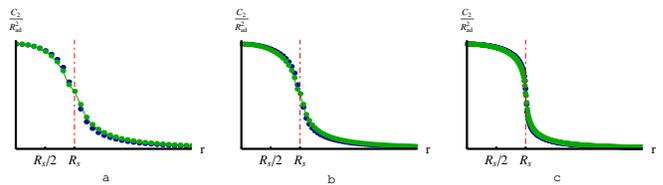}
\caption{The normalization factor, $N_{\mbox{\scriptsize tot}, (2)}$ in 2d using numerical integration for three different values of cut-offs $r_c/R_s=0.1,0.05,0.01$ ((a)-(c)). The red dashed line indicates the radius of the swarm at $r=R_s$. The green points are calculated from the lattice model approximation for comparison.}
\label{NtotDensities}
\end{figure}

\begin{figure}[tb]
\centering
\includegraphics[width=\linewidth]{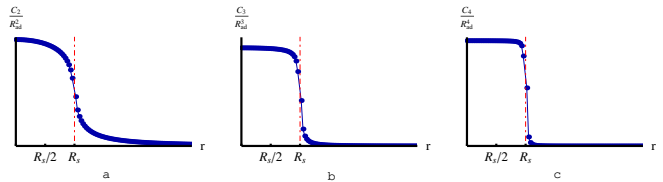}
\caption{The normalization factor, $N_{\mbox{\scriptsize tot}, (n)}$ in 2d for three different values of $n$ with the same value of cut-off  $r_c/R_s=0.03$ (a) - $n=2$ in units of $C_{2}/R_{\mbox{\scriptsize ad}}^2$ (b) - $n=3$ in units of $C_{3}/R_{\mbox{\scriptsize ad}}^3$ (c) - $n=4$ in units of $C_{4}/R_{\mbox{\scriptsize ad}}^4$. The red dashed line indicates the radius of the swarm at $r=R_s$.}
\label{Ntotn}
\end{figure}

To avoid a divergence of  $N_{\mbox{\scriptsize tot}, (n)}$ at $y_0=R_{s}$ we maintain a cutoff distance to the edge $R_{s}\pm r_c$. In Fig.~\ref{NtotDensities} we give $N_{\mbox{\scriptsize tot}, (2)}$ (Eqs.~(\ref{RegularizedNtot2d}),(\ref{RegularizedNtot3d})) for $y_0<R_{s}$, and for $y_0\geq R_{s}$ (Eq.~(\ref{Ntot2d})), with three different values of cut-offs $r_c/R_s=0.1,0.05,0.01$. The results agree with a lattice model where the size of the cut-off defines the lattice spacing. The normalization factor $N_{\mbox{\scriptsize tot}, (n)}$ for different values of $n$ is given in Fig.~\ref{Ntotn}. These figures show that for a discrete swarm, with a minimal separation between the organisms $r_c$, both the stimulus and the total amplitude are well defined throughout the domain.

\begin{figure}[tb]
\centering
\includegraphics[width=\linewidth]{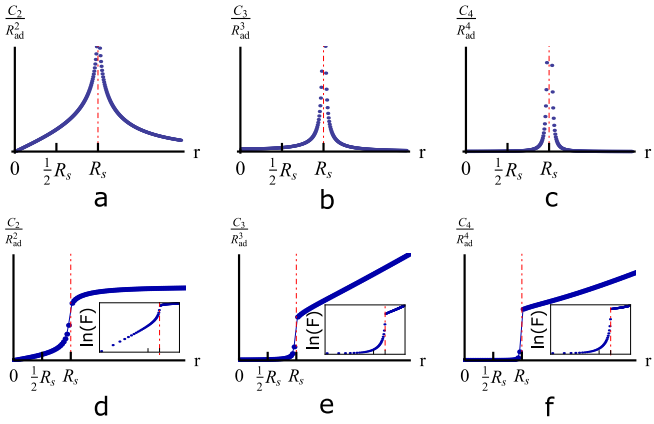}
\caption{
The total stimulus in 2d as a function of the distance from the center of the disk $r$ (from the numerical integration) (a) $n=2$ in units of $C_{2}/R_{\mbox{\scriptsize ad}}^2$ (b)  $n=3$ in units of $C_{3}/R_{\mbox{\scriptsize ad}}^3$ (c) $n=4$ in units of $C_{4}/R_{\mbox{\scriptsize ad}}^4$. The effective force in 2d as a function of $y_0$ (d)  $n=2$ in units of $C_{2}/R_{\mbox{\scriptsize ad}}^2$ (e) $n=3$ in units of $C_{3}/R_{\mbox{\scriptsize ad}}^3$ (f) $n=4$ in units of $C_{4}/R_{\mbox{\scriptsize ad}}^4$. The red dashed line indicates the radius of the swarm at $r=R_s$. (A cut-off value of $r_c/R_s=0.03$ is used). The insets show the same plots in log-log scale.}
\label{Feff2d}
\end{figure}

The continuum integration for $S_{(n)}$ is done numerically for various values of $n$ (Fig.~\ref{Feff2d}a-c). The continuum integral is well behaved outside the disk. However it seems to be divergent for $y_0<R_s$, where this apparent divergence is a manifestation of the fact that the force diverges with a higher power than the mass contribution, which increases as the area $r^2$. We therefore introduce a cut-off ($r_c$) around the calculated point (Fig.~\ref{coordinates2}b). Using the cut-off we find that $S_{(n)}$ does not diverge inside the disk, for the following reason: For any point inside the disk there are sources that give divergent contributions in all directions and eventually cancel out to give a finite stimulus in the limit of $r_c\rightarrow0$. On the boundary, however, all the sources are on one side and they add up to give a divergence, as shown in Fig.~\ref{Feff2d}a-c. When considering the finite distance between organisms in the swarm, of order $r_c=\sigma^{-1/2}$, the stimulus near the swarm edge ($y_0=R_s\pm r_c$) is also finite.


Next we calculate the total amplitude factor, $N_{\mbox{\scriptsize tot},(n)}$. We find, as for the stimulus, that the continuum description is well behaved outside the disk (Eq.~\ref{Ntot2d}). However, inside the disk we now have a real divergence for all the points, in the limit where $r_c\rightarrow0$. The reason is that unlike the vectorial sum appearing in $S_{(n)}$, where contributions from different directions cancel, we sum absolute amplitudes which add up. We therefore need to carry out the numerical integrations with a particular choice of $r_c$.

We obtain the effective force by dividing the total stimulus by the normalization factor (Eq.~(\ref{perfect})), as shown in Fig.~\ref{Feff2d}d-f for different values of $n$, using a cut-off value of $r_c/R_s=0.03$. Comparing the effective force to the stimulus, we see that perfect adaptivity removes the decay of the effective force outside the disk. This is true below a distance of $N^{\frac{1}{n}}\,R_{\mbox{\scriptsize ad}}$ (Eq.~(\ref{Fadapt})) from the edge of the disk. At large distances $r-R_s\gg N^{\frac{1}{n}}R_{\mbox{\scriptsize ad}}$ adaptivity is ineffective, and the effective force becomes identical to the stimulus (see Fig.~\ref{theforce}a).

Inside the disk, we find that functionally the stimulus and the effective force are very similar. However, they have very different dependencies on the density of sources. This can be clearly shown by calculating the force near the center of the swarm ($y_0 \ll R_s$). We get from Eq.~(\ref{TotalStimulus2d})
\begin{equation}
S_{(n)}(y_0)=\frac{\pi\,C_n\,\sigma y_0}{R_{s}^{n-1}}+\mathcal{O}(y_0^2),
\label{Sn2dcenter}
\end{equation}
and from Eqs.~(\ref{RegularizedNtot2d},\ref{RegularizedNtot3d}) (assuming $r_c<R_s$)
\be
N_{\mbox{\scriptsize tot}, (n)}=\begin{cases}
   2\,\pi\,\sigma\,C_2\ln(R_s\,\sqrt{\sigma})+\mathcal{O}(y_0), & n=2\\
    \\
    \frac{2\,\pi\,C_n\,\sigma}{(n-2)}\(\sigma^{\frac{n-2}{2}}-\frac{1}{R_s^{n-2}}\)+\mathcal{O}(y_0), & n>2.
  \end{cases}
\ee

Therefore the effective force at the center is
\be
F_{(n)}(y_0)=\begin{cases}
\frac{C_2}{R_{\mbox{\scriptsize ad}}^2}\cdot\frac{y_0}{2\,R_s \ln(R_s\sqrt{\sigma})}+\mathcal{O}(y_0^2), & n=2\\
\\
\frac{C_n}{R_{\mbox{\scriptsize ad}}^n}\cdot\frac{(n-2)y_0}{2\,R_s\left[R_s^{n-2}\,\sigma^{\frac{n-2}{2}}-1\right]}+\mathcal{O}(y_0^2). & n>2
 \end{cases}
 \label{F2dcenter}
\ee

Both the stimulus and the effective force are linear with the displacement $y_0$ from the disk center, for all values of $n$. However, while $S_{(n)}$ is linear in $\sigma$ (Eq.~(\ref{Sn2dcenter})), the effective force is a \textit{decreasing} function of $\sigma$ (Eq.~(\ref{F2dcenter})). This behavior, for all $n\ge2$, is due to the fact that the intensity increases with increasing density faster than the geometric decrease in the number of neighbors.



\section{The divergence of $N_{\mbox{\scriptsize tot}}$ for $n=2$ and $n=4$ in two dimensions (cartesian coordinates)}

 \label{AppG}

 In order to show the divergence in cartesian coordinates let us look at the hypergeometric functions in Eq.~(\ref{Ntot2d}) and start from $n=2$.
Since
\[F^{\pm}_{2}(\tx)=\frac{|\tx|\,\arctan\left(\frac{\ty0\pm\sqrt{1-\tx^2}}{\tx}\right)}{|\ty0\pm\sqrt{1-\tx^2}|}, \]
at $\tx=0$ we have the following expansion:
\be
F^{\pm}_{2}(\tx)=\frac{\pi}{2|\ty0\pm1|}|\tx|+\mathcal{O}(\tx^2).
\ee
Substituting into Eq.~(\ref{Ntot2d}) and taking the limit $\tx \rightarrow 0$ we see that
\[N_{\mbox{\scriptsize tot}, (n)} \sim \int_{-1}^{1}\frac{\pi d\tx}{|\tx|}\] for $|\ty0| \leq 1$ and
\[N_{\mbox{\scriptsize tot}, (n)} \sim \mathcal{O}(1)\] for  $|\ty0| \geq 1$.

We can find the expression for $F^{\pm}_{4}(\tx)$ using the recursive formula (see for instance~\cite{wang1989special}):
\begin{widetext}
\be
(\beta-\gamma)\, _{2}\!F_{1}(\alpha,\beta-1,\gamma;z)+\(\gamma-2\,\beta+(\beta-\alpha)z\)\, _{2}\!F_{1}(\alpha,\beta,\gamma;z)=\beta\,(z-1)\, _{2}\!F_{1}(\alpha,\beta+1,\gamma;z),
\ee
\end{widetext}
which gives
\be
_{2}\!F_{1}(1,-\frac{1}{2},\frac{3}{2};z)=\frac{1-z}{2}\left[_{2}\!F_{1}(1,\frac{1}{2},\frac{3}{2};z)+_{2}\!F_{1}(1,\frac{3}{2},\frac{3}{2};z)\right],
\ee
and since $F_{1}(1,\frac{3}{2},\frac{3}{2};z)=1/(1-z)$ we get:
\be
F^{\pm}_{4}(\tx)=\frac{\tx^2+(\ty0\pm\sqrt{1-\tx^2})^2}{2\,\tx^2}F^{\pm}_{2}(\tx)+\frac{1}{2}.
\ee
Substituting into Eq.~(\ref{Ntot2d}) we see that in the limit $\tx \rightarrow 0$
\[N_{\mbox{\scriptsize tot}, (n)} \sim \int_{-1}^{1}\frac{\pi d\tx}{|\tx|}\] for $|\ty0| \leq 1$ and
\[N_{\mbox{\scriptsize tot}, (n)} \sim \mathcal{O}(1)\] for  $|\ty0| \geq 1$.

In a similar way one can derive a similar result for $n=6$.

\section{Jeans Instability (rigorous derivation)}

 \label{AppH}

Let $f(\vec{r},\vec{v},t)$ be the phase-space density of organisms. $f(\vec{r},\vec{v},t)d^{3}\vec{r}d^{3}\vec{v}$ is the number of organisms having positions in the small volume $d^{3}\vec{r}$ centered on $\vec{r}$ and velocities in the small range $d^{3}\vec{v}$ centered on $\vec{v}$. The collisionless Boltzman equation for the collection of organisms is
\be \label{boltzman}
\frac{\partial f}{\partial t}+\vec{v}\cdot \vec{\nabla} f+\vec{F}_{(n)}\cdot\vec{\nabla}_{\vec{v}}f=0,
\ee
where $\vec{F}_{(n)}$ is the effective force on an organism. Let us consider a small fluctuation in the density function
\be
f(\vec{r},\vec{v},t)=f_0(\vec{r},\vec{v},t)+\epsilon f_{1}(\vec{r},\vec{v},t). \label{dist1st}
\ee
where $f_0(\vec{r},\vec{v},t)$ is the phase-space density of the background and $f_{1}(\vec{r},\vec{v},t)$ is the small fluctuation. The number density of the background is
\be
\rho_{0}(\vec{r})=\int d^{3} \vec{v}\,f_{0}(\vec{r},\vec{v},t),
\ee
and the small fluctuation in the number density is
\be
\rho_{1}(\vec{r})=\int d^{3} \vec{v}\,f_{1}(\vec{r},\vec{v},t).
\ee
The force in the total adaptive regime is given in Eq.~(\ref{perfect}):
\be
\vec{F}_{(n)}=\frac{|C_n|\vec{S}_{(n)}}{R_{\mbox{\scriptsize ad}}^{n}\,N_{\mbox{\scriptsize tot},(n)}}.
\ee

In order to represent the small fluctuation in density we expand $\rho(\vec{r})$ as described above and the stimulus correspondingly,
\be
\vec{S}_{(n)}=\vec{S}^{(0)}_{(n)}+\epsilon\vec{S}^{(1)}_{(n)}.
\ee
 We construct a fictitious homogeneous equilibrium by taking $\vec{S}^{(0)}_{(n)}\equiv 0$ and $\rho_0(\vec{r})=constant$. This is the so-called Jeans swindle. There is no formal justification for this construction but it gives us the instability in the first order perturbation. Therefore we can write to first order in epsilon:
\be \label{force_1st}
 \vec{F}_{(n)}^{(1)}=\epsilon\frac{|C_n|\vec{S}^{(1)}_{(n)}}{R_{\mbox{\scriptsize ad}}^{n}\,N_{\mbox{\scriptsize tot},(n)}},
\ee
where
\be
N_{\mbox{\scriptsize tot},(n)}= |C_n|\,\rho_0\,\int\frac{d^{3}\vec{r'}}{\mid\vec{r}-\vec{r'}\mid^{n}}
\ee
is the total amplitude. Since we assume an infinite background with uniform density, the force at the center of a large uniform sphere in Eq.~(\ref{sn3dcenter})
will give us a good approximation to the perturbative force:
\be
\vec{S}^{(1)}_{(n)}=\frac{4\,\pi\,\rho_{1}\,C_n\,r}{3\,R_{s}^{n-2}}+\mathcal{O}(\frac{r^3}{R_{s}^{3}}),
\ee
when $r$ is the distance from the center (and $C_n<0$ for an attractive force):
\be
\vec{\nabla}\cdot\vec{S}^{(1)}_{(n)}\sim\frac{4\,\pi\,\rho_{1}\,C_n}{R_{s}^{n-2}}. \label{J1}
\ee

The gravitational force can be expressed as a gradient of a potential:
\be
\vec{S}_{(n)}^{(1)}=-\vec{\nabla}\Phi_1(\vec{r},t).
\ee
Substituting Eqs. (\ref{force_1st}) and (\ref{dist1st}) into Eq. (\ref{boltzman}) we get
\be \label{J2}
\frac{\partial f_1}{\partial t}+\vec{v}\cdot\vec{\nabla}f_1+\frac{|C_n|}{R_{\mbox{\scriptsize ad}}^{(n)}\,N_{\mbox{\scriptsize tot},(n)}}\,\vec{S}_{(n)}^{(1)}\cdot\vec{\nabla}_{\vec{v}}f_0=0.
\ee
We substitute a wave-like periodic solution for the disturbance in the phase space distribution and the potential (Eqs. (\ref{J1})-(\ref{J2})):
\bea
f_{1}(\vec{r},\vec{v},t)&=&f_{a}(\vec{v})e^{i(\vec{k}\cdot\vec{r}-\omega t)}\nonumber\\
\Phi_1(\vec{r},t)&=&\Phi_{a}e^{i(\vec{k}\cdot\vec{r}-\omega t)},
\eea
and get the following dispersion relation for $\omega(\vec{k})$:
\be
1-\frac{4\,\pi\,C_n}{k^2\,R_{s}^{n-2}}\cdot\frac{|C_n|}{R_{\mbox{\scriptsize ad}}^n\,N_{\mbox{\scriptsize tot},(n)}}\int\frac{\vec{k}\cdot\frac{\partial f_0}{\partial \vec{v}}d^3 \vec{v}}{\vec{k}\cdot \vec{v}-\omega}=0.
\ee
Assuming a gaussian distribution of velocities
\be
f_{0}(\vec{v})=\frac{\rho_0}{(2\,\pi\,\sigma_v^2)^{\frac{3}{2}}}\,e^{-\frac{v^2}{2\,\sigma_v^2}},
\ee
where $\sigma_v$ is the width of the gaussian and choosing the $v_x$ direction to lie in the direction of $\vec{k}$ we get
\be
1+\frac{2\,\sqrt{2\,\pi}\,C_n\,|C_n|\,\rho_0}{R_{\mbox{\scriptsize ad}}^n\,R_{s}^{n-2}\,N_{\mbox{\scriptsize tot},(n)}\,k\,\sigma_v^3}\,\int_{-\infty}^{\infty}\frac{v_x\,e^{-\frac{v_x^2}{2\,\sigma_v^2}}}{k\,v_x-\omega}dv_x=0. \label{dis_spe}
\ee
From Eq. (\ref{dis_spe}) we can write the dispersion relation $\omega(k)$ explicitly. The
The critical $k$ is obtained when $\omega=0$:
\be
k_{\mbox{\scriptsize Jeans}}^2=\frac{4\,\pi\,C_{n}^{2}\,\rho_0}{R_{\mbox{\scriptsize ad}}^n\,R_{s}^{n-2}\,N_{\mbox{\scriptsize tot},(n)}\langle v^2 \rangle},
\ee
since it can be shown that for $k<k_{\mbox{\scriptsize Jeans}}$ the modes are unstable, namely $\omega^{2}<0$ (see~\cite{binney2011galactic}, p. 292).
Subsitituting $N_{\mbox{\scriptsize tot},(n)}$ from Eqs.~(\ref{Ntot2},\ref{Ntotcenter}) and taking $R_s\sim L$ and $k \sim 2\,\pi/L$, we obtain a similar criteria for instability, up to a multiplicative factor, as in Eqs.~(\ref{crit1})-(\ref{criticalDens}).

\bibliography{acoustics_1}
\end{document}